# Practical Aspects of the Bitcoin System

Artus Krohn-Grimberghe, University of Paderborn, Germany

Christoph Sorge, University of Paderborn, Germany

## Abstract

Digital payment schemes show an ever increasing importance. Out of the countless different schemes available this article focuses on the popular Bitcoin system. The authors provide a description of Bitcoin's unique technological basis and its accompanying ecosystem of users, miners, trading platforms and vendors. Furthermore, this article discusses Bitcoin's currency-like features and the first regulatory actions take in the European Union and in the United States of America.

## 1. Introduction[1]

Though internet-based business has become an important factor for the economy, there is still no widely accepted payment scheme that can be considered equivalent to cash. Most systems rely on a central payment provider, which might be a credit card provider or the operator of its own payment scheme. The payment provider has to be financed, so it usually charges a transaction fee. In addition, most payment systems are not anonymous, so the payment provider can track all transactions of its users, or decide to block payments by certain users. Payment schemes offering full anonymity exist, but they make up a small percentage of the market only. Researchers have come up with numerous protocols for anonymous online payment, most of which also require existence of a central server.

In November 2008, someone using the pseudonym Satoshi Nakamoto posted his idea for a completely decentralized electronic currency called "Bitcoin" to a cryptography mailing list. The system was actually launched in 2009 and has gained popularity since then. In addition to online retailers accepting the currency, a number of other players offer services for the Bitcoin system. However, the adoption of the Bitcoin system is still in an early stage. Attacks on Bitcoin service providers threaten to affect user acceptance, and doubt has been cast on their professionalism. How should agencies in charge of financial market supervision react to the developments? Our goal is to answer this question, while (for the most part) abstracting from the details of legislation in any specific country.

In this article, we present the technical basis of Bitcoin (Section 2) and discuss its legal classification (Section 3). We then report on developments in a number of countries (Section 4) alongside with major incidents in the Bitcoin ecosystem (Section 5). We cope with data protection issues (Section 6) before concluding the article in Section 7.

## 2. Bitcoin: The Technology

In this section, we describe the main ideas of the Bitcoin system while leaving out details and optimizations that are not essential to the concept.

---

[1] The authors have previously published an article about the legal classification of Bitcoin according to German law (Sorge & Krohn-Grimberghe, 2012).

Bitcoin was designed as a completely decentralized "electronic cash" system, though the system lacks the anonymity properties typically expected from electronic cash (at least by cryptographers), such as unlinkability of different transactions, and user anonymity against collaborating transaction partners. Instead of the concept of electronic coins that most e-cash schemes use, Bitcoin is focused around transactions[2]. Any participant can create an arbitrary number of accounts, and "electronic money" can be transferred from one account (or even several accounts at the same time) to other accounts.

Technically, an account is represented by a key pair: The private key is used to sign transactions using that account as a source, while these signatures can be verified with the corresponding public key. The public key can also be seen as an account number, i.e. as an identifier for the account, and its hash is generally known as a Bitcoin address. To verify if a transaction is valid, one has to check whether a sufficient amount of money was transferred to that account beforehand. The sender has to provide references to incoming transactions to his accounts which have not been used yet.

Unfortunately, the system described so far does not prevent double spending. As there is no central entity, double spending prevention is achieved by publishing all transactions, thus allowing any participant to verify the validity of a transaction. This way, if someone attempted spending money a second time, anyone could see the previous transaction and therefore know that the second one is invalid.

The lack of a central, trusted authority, however, also leads to another problem: A fake transaction history cannot be distinguished from a real one, as there is no trust anchor available for authentication. Bitcoin deals with this issue by making it computationally difficult to compute a valid transaction history. Transactions are combined into blocks; participants that have checked the transactions and want to confirm them ("miners") have to compute a proof of work that uses as input both the new block and (conceptually) the complete transaction history before that block. The proof of work is based on a cryptographic hash function: The miners must find a value that, together with the inputs, yields a hash value with a certain property. Including the previous transaction history means that with each added block, the confirmation of previous blocks becomes stronger. The longest transaction history, called the blockchain, is considered authoritative, and will be used by all miners as a basis for adding additional transaction blocks.

A miner that has completed a proof of work collects all transaction fees associated with the processed transaction (just like a credit card issuer) as well as a "block reward", which is newly generated "money"—both is not guaranteed by a technical mechanism, but by consensus of participants in the Bitcoin system to accept these rewards for the miners. Over time, proofs of work become more difficult and block rewards become smaller—therefore, the creation of new "money" is limited. The idea behind the system is that due to this limited availability, the value of each unit increases, while its divisibility makes sure the system will not fail due to lack of "coins". Also, potentially increasing transaction fees are supposed to provide an incentive to miners even in the presence of diminishing block rewards.

---

[2] The term "coin" is used in the original Bitcoin description, but not in the usual sense of objects with a fixed value.

There are two main concerns about the Bitcoin protocol from a technical perspective. Firstly, there is serious doubt about its scalability. At least all miners have to be informed about all transactions. A possible reaction is to limit the number of miners, though this goes against the peer-to-peer paradigm. Even now, most Bitcoin users do not mine themselves. Most of the computational power is actually concentrated in a few "mining pools", which leads to the second problem: An attacker that has more computational power available than all honest miners combined could create and confirm bogus transactions—since the attacker's version of the blockchain would grow faster than the correct one, it would quickly be accepted as legitimate. Given that, at the time of this writing, the largest mining pool had more than 20% of the combined computational power (expressed in GHash/s) available, this attack hardly seems impractical. Additionally, Bitcoin transactions take a considerable amount of time (minutes to hours) depending on the amount of confirmations required. This drawback is countered by the creating of whitelisted addresses, so called "green addresses"[3] which are trusted not to double spend. However, for transactions involving those whitelisted addresses, Bitcoin is no longer a peer to peer system.

## 3. Bitcoin: A Currency?

From a legal perspective, the question arises what Bitcoin is, and if/how the system should be regulated. We start with some general considerations, before having a closer look at the legal situation in Europe and the United States of America. The Bitcoin Wiki[4] states that Bitcoin is a "digital currency". However, this term only applies in its colloquial use. The use of the term "currency" by economists (and, consequently, in jurisprudence) is limited to state-approved money. Some authors define currency to only include "paper money and coins" (Mishkin, 2004) (Campbell & Campbell, 1988). Following this definition, the term "digital" currency becomes meaningless.

### 3.1. Electronic money and money

In the member states of the European Union, Directive 2009/110/EC of the European Parliament and of the Council defines the concept of "electronic money", which, at first glance, seems to characterize Bitcoin more appropriately.

According to Article 2 of the directive, electronic money "means electronically, including magnetically, stored monetary value as represented by a claim on the issuer which is issued on receipt of funds for the purpose of making payment transactions […], and which is accepted by a natural or legal person other than the electronic money issuer".

To fulfill this definition, the Bitcoin system would have to use an "electronically stored monetary value". In fact, a Bitcoin client stores key pairs representing accounts. Account balances do not correspond to a fixed value in any "outside" currency, but have meaning only within the system[5]. However, this does not exclude the representation of a "monetary value". Key pairs also do not represent electronic coins, as the value associated with a key pair is only determined by keeping track of previous transactions; in this respect, Bitcoin is comparable to book money. We still consider Bitcoin to use an "electronically stored monetary value": The value of a user's "accounts" can be easily computed using the stored information, and from the user's perspective, there is no major difference to using other electronic payment systems.

---

[3] http://bitcoin.stackexchange.com/questions/1730/what-are-green-addresses
[4] https://en.bitcoin.it/w/index.php?title=Main_Page&oldid=35321
[5] In practice, Bitcoins can be exchanged to Dollars or Euros at varying exchange rates. This is not a property of the system, but a feature offered by service providers.

To qualify as electronic money, however, there would have to be "a claim on the issuer". The European legislator was obviously thinking of payment systems operated by a single issuer. While Bitcoin is a decentralized system, there are entities that might be thought of as "issuers": Miners get a block reward after having completed a proof of work, and by that, they generate new Bitcoins. The role of a miner spending Bitcoins, however, is not much different from the role of any other participant spending Bitcoins. We could, of course, consider all participants of the Bitcoin system as "issuers". Still, even if we do, there is neither is issuance "on receipt of funds" nor a tangible creator of a Bitcoin block. Most importantly, there is also no "claim on the issuer", i.e. no legal requirement for an "issuer" to pay out a corresponding amount of money. Bitcoin is not linked with any traditional money format. The value of a key pair associated with a certain amount of Bitcoins only lies in the expectation that others will accept outgoing transactions from that key pair, e.g. in exchange for goods or services rendered.

As a consequence, Bitcoin does not fulfill the definition of electronic money in the European Union.

Next, we discuss which properties of *money* are shared by Bitcoin. There is no generally accepted definition of the term (Proctor, 2005, margin number 1.08). Its usage differs between different fields of law (e.g. between penal provisions regarding the counterfeit of money, and banking regulations).

Many of these definitions, for example in criminal law, require money to be issued by the state or a state-authorized agency; this is obviously not the case for Bitcoin. Economic definitions of money require its widespread acceptance. For example, Mishkin (Mishkin, 2004) defines money as "anything that is generally accepted in payment for goods or services or in the repayment of debts" (Proctor's definition (Proctor, 2005) also contains this requirement). Bitcoin is not "generally accepted"; the lack of widespread acceptance has also been a reason for the United Kingdom's Financial Services Authority and the Swedish Finansinspektionen to classify Bitcoin as not being money[6].

Even though Bitcoin does not qualify as money according to most definitions, we take a look at the functions of money from an economic perspective (Mishkin, 2004) which are also shared by the European Central Bank (European Central Bank, 2012) and the German Bundesbank (Bundesbank, 2013):

- Money can be used as a store of value. Acquired Bitcoins do not have to be spent immediately; in principle, the key pairs can be stored for years before the value is retrieved. The value of Bitcoins changes over time. The same is true for conventional currencies (though fluctuations are typically less extreme in that case). Barring hyperinflation, fluctuations of value do not prevent fulfillment of the store of value function.
- Money serves as a medium of exchange. Goods or services can be exchanged for Bitcoins, instead of exchanging goods directly (e.g. trading pears for apples). We discuss this property in more detail in Section 3.2.
- Finally, money functions as a unit of account. A more detailed discussion of this property follows in Section 3.3.

---

[6] Personal communication.

## 3.2. Medium of Exchange

The function of money as a "medium of exchange" describes its use in trade to avoid the use of a direct barter system.

Proctor (Proctor, 2005) cites the description from the case of Moss v. Hancock as the perhaps best known British judical definition of money as a medium of exchange: money is "that which passes freely from hand to hand troughout the community in final discharge of debts and full payment of commodities, being accpeted equally without reference to the character or credit of the person who offers it and without the intention of the person who receives it to consume it or apply it to any other use than in turn to tender it to others in discharge or debts or payment for commodities."

The "discharge of debt" is certainly possible with Bitcoin, as a debtee is free to accept Bitcoin – though there is no obligation to do so. The relevant question is whether Bitcoin is actually used for the purpose, i.e. in payment of commodities – even if it is not legal tender. As there are merchants accepting Bitcoin, and the Bitcoin system was even designed for that purpose, we conclude that Bitcoin can fulfill the "medium of exchange" function. This view is shared by the Swedish Finansinspektionen, which considers Bitcoin as a (regulated) "means of payment" since late 2012, and the European Central Bank.

However, the actual use of Bitcoin as a medium of exchange is very limited as of mid 2013. This lack of actual use is the reason for the British Financial Services Authority not to consider Bitcoin as money.

## 3.3. Unit of Account

There is no legal definition of the term "unit of account", but the function of a unit of account is clear in economic literature (e.g. (Mishkin, 2004)): The prices of goods and services can be expressed, or in other words, their value can be measured using the unit of account. Expressing values using the unit "Bitcoin" is very uncommon. Even if online retailers accept Bitcoin, prices are usually stated in US Dollars, and an exchange rate is applied when an actual payment is made. In principle, Bitcoin could be used as a unit of account: this is true for any good. The good does not even have to be available or manageable, as long as the relation of its value to the value of other goods can be determined. Trying to find a proper delineation of the term "unit of account", we consider the example of Special Drawing Rights (SDRs), defined by the International Monetary Fund based on the value of several currencies. They are unanimously considered as a unit of account. The only distinction between SDRs and arbitrary goods, like a kilogram of wheat, is the *intended* and the *actual* use: SDRs have the specific purpose of being used as a unit of account, i.e. to express the value of some other goods. While SDRs only play a role in a very narrow sector, they are also actually used for that purpose. The situation is similar for Bitcoin. While the original Bitcoin article by Nakamoto focuses on the technical aspects of an electronic payment system, the fact that Bitcoin constitutes a unit of account is inherent in its design: As new Bitcoins can be created by investing computational power, the system cannot just be used to carry out payments in any existing currency. Moreover, speculation with Bitcoin (exploiting varying exchange rates with currencies like US Dollar or Euro) is actually taking place and takes advantage of the fact that Bitcoin is an independent unit of account. We conclude that Bitcoin fulfills this function.

To summarize, Bitcoin does not constitute electronic money in the sense of the Directive 2009/110/EC of the European Parliament and of the Council. We find that Bitcoin has a potential to fulfill all the defined roles of money in theory, but lacks the widespread acceptance of actual money.

## 4. Regulatory action

Over the course of the last two and a half years Bitcoin has received a growing amount of attention from regulatory bodies. This section tries to tie together what is already known.

On the European level, there is no regulation of Bitcoin as of June 2013. In October, 2012, however, the European Central Bank (ECB) published a report on "Virtual Currency Schemes", trying to establish a notion of "virtual money" or "virtual currency" which essentially "act as medium of exchange and unit of account within a particular virtual community" (European Central Bank, 2012). Key aspects to the ECB's definition of virtual currencies are the (solely) digital and a lack of regulation. The report clearly states that the difference between electronic money schemes and virtual money schemes lies in "the currency being used as the unit of account [having] no physical counterpart with legal tender status." There are no legal consequences of the classification as virtual money, as the term has not been used in legislation. Moreover, using the lack of regulation as part of the definition does not help to decide whether Bitcoin should be regulated.

Possibly the first country to regulate the Bitcoin market was Germany. The German Bundesanstalt für Finanzdienstleistungsaufsicht (Federal Agency for Financial Market Supervision) together with the German Bundesbank and the German Ministry of Finance concluded that Bitcoin is a "unit of account" in mid September, 2011 – this result is in line with our assessment (Section 3.3). German law considers a unit of account as a financial instrument[7]. As a consequence, certain services related to Bitcoin, including the operation of a multilateral trading system, require permission from the German Federal Agency for Financial Market Supervision (BaFin). The authors know of at least one instance where the operation of a Bitcoin trading facility in Germany was suspended due to this regulation.

The next regulatory action the authors are aware of has taken place in Sweden. While during the mid-2011 till mid-2012 time frame the Swedish Finasinspektionen (Financial Services Inspection) did not regard Bitcoin as a regulated matter, this view changed during late summer or fall 2012. Since then, Bitcoin is considered a "means of payment" (medium of exchange) in Sweden. The effect of this is that anyone in Sweden who intends to facilitate a market for Bitcoin has to register with Finansinspektionen and fulfill the requirements on financial institutions.

Official information on the regulation of Bitcoin in the United States of America did not appear before March, 2013, when the United States Financial Crimes Enforcement Network (FinCEN) issued a guidance regarding "virtual currencies" and their relation to money services business (the regulated transmission or conversion of money in the United States of America). According to this guidance, exchanging Bitcoins and mining Bitcoins for profit may be a regulated activity in the United States of America (Financial Crimes Enforcement Network, 2013). Furthermore, at least since May 2013 the Commodity Futures Trading Commission (CFTC) is exploring regulation for Bitcoin, too. Later in the same month Mt Gox' account at the Dwolla payment network was seized for unlicensed money transmission by a Baltimore judge (Ars Technica, 2013).

---

[7] Gesetz über das Kreditwesen, Section1, subsection11

To the authors' knowledge, no other countries have taken regulatory action as of May 2013. This includes Japan, home to the largest Bitcoin exchange, MtGox.

As consumer protection is a major motivation for regulation of financial markets, we have a look at incidents that could be seen as a justification for the necessity of regulatory action.

## 5. Incidents

Although information posted in the Bitcoin forums has to be taken with a grain of salt, it is quite clear that till May 2013 Bitcoins worth more than $3.000.000 USD (priced at their MtGox conversion rate at the time of the incident)[8] have been stolen or lost, with the dark figure probably being much higher. Most cases involved acts of negligence (unencrypted Bitcoin wallets containing hundreds or thousands of Bitcoins[9], not having backups of the private keys for very large Bitcoin wallets[10], simple exploitability of web apis[11][12][13] and security concepts[14][15], scam[16][17], pyramid schemes[18] and therelike[19]). Additionally, much more money is being lost due to trade manipulation including schemes as moving bid / ask walls and distributed denial of service attacks against brittle match making systems; the latter make it impossible to sell during flash crashes, causing short squeezes[20]. The current Bitcoin ecosystem is still very immature with inadequate infrastructure and inapt key players, leaving the impression of a Wild West state. The impressive track record of incidents in the Bitcoin ecosystem combined with the increasing financial volume indicates that further regulation and thus more professional players in the key roles might be required.

---

[8] https://bitcointalk.org/index.php?topic=83794.0
[9] https://bitcointalk.org/index.php?topic=16457.msg214423#msg214423; User "allinvain" lost 25.000 Bitcoins when someone stole his unencrypted Bitcoin wallet and many other postings like this.
[10] https://support.mtgox.com/entries/20357051-mt-gox-the-world-s-largest-bitcoin-exchange-to-acquire-bitomat-pl-compensate-loss-of-bitcoins Bitomat lost 17.000 Bitcoins due to not having backups when the virtual machine running the exchange incidentially got deleted.
[11] https://bitcointalk.org/index.php?topic=66979.0 Bitcoinica lost 43.500 Bitcoins when their web hoster Linode was attacked in March 2012
[12] https://bitcointalk.org/index.php?topic=66916.0 The mining pool Slush lost 3.000 Bitcoins in the Linode hack
[13] http://btcbase.com/2012/05/14/einbruch-bei-bitcoinica-com-uber-18000-bitcoins-entwendet/ Bitcoinica lost 18.000 Bitcoins and their master password in May 2012
[14] http://bitcoinmagazine.com/bitcoinica-stolen-from-again/ Bitcoinica lost another 40.000 USD and 40.000 Bitcoins due to the lost password still being in use in July 2012.
[15] http://www.heise.de/newsticker/meldung/Bankraub-und-Erpressung-mit-Bitcoins-1702157.html Bitfloor lost 24.000 Bitcoins due to unencrypted backups in September 2012
[16] http://www.betabeat.com/2011/08/05/mybitcoin-disappeared-with-bitcoins/ Online wallet service MyBitcoin disappeared together with more than 25.000 Bitcoins stored there.
[17] https://bitcointalk.org/index.php?topic=65867 The World Bitcoin Exchange operator disappeared together with 25.000 AUD; the 1.769 Bitcoins could be recovered, however.
[18] https://bitcointalk.org/index.php?topic=102079.160 Bitcoin Savings & Trust managed by PirateAt40 promised returns of 1% per day. He escaped with at least 10.000 Bitcoins.
[19] https://bitcointalk.org/index.php?topic=132070.0 Nearly 20.000 Bitcoins held as customer funds by Bitcoin exchange BitMarket.eu were lost when used for speculation.
[20] Personal communication with a professional Bitcoin trader.

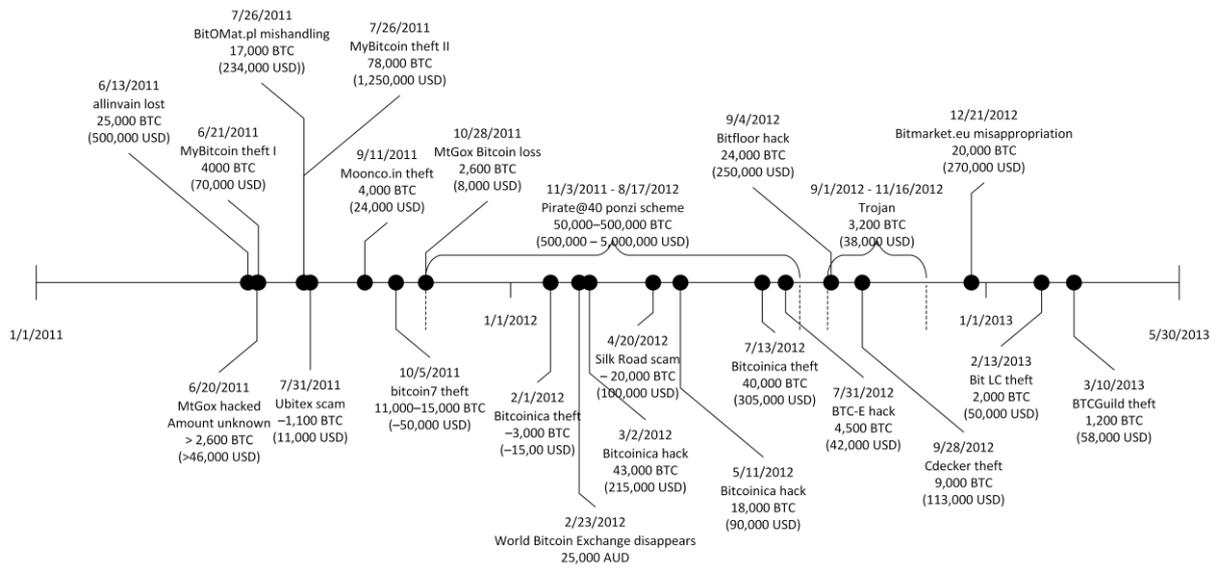

## 6. Data protection

When dealing with Bitcoin, financial market regulation is the most obvious legal aspect to be considered. However, Bitcoin is also a Peer-to-Peer system processing transaction data, which might be sensitive – and thus legally protected – as well. A number of publications have discussed the potential of de-anonymizing Bitcoin transactions. Attacks try to merge different transactions and/or key pairs related to the same user (Reid & Harrigan, 2013), (Ron & Shamir, 2012), or they are directly based on the underlying Peer-to-Peer network[21]. As a consequence, transaction data might be considered as personal data – especially when combined with IP addresses of participants. In the European Union, the processing of personal data is regulated in the European Data Protection Directive (Directive 95/46/EC). The directive does not prevent operation of the Bitcoin system itself, but usage of collected data for other purposes may be problematic. The legal status of Peer-to-Peer systems with respect to data protection legislation is still unclear, but given the lack of other options, each participant is likely to be considered as a "controller", i.e. as the entity responsible for adherence to data protection legislation. The problem disappears when using the Zerocoin approach (Miers, Garman, Green, & Rubin, 2013), an extension to Bitcoin that achieves privacy of payments.

## 7. Summary

In this article, we have discussed the Bitcoin system from a legal and regulatory perspective. We have shown that Bicoin is capable of fulfilling all functions of money, but cannot currently be considered as money, as it is not widely accepted as such. However, authorities in Germany, Sweden and the United States of America have initiated regulation of the Bitcoin market based on the classification on Bitcoin as a unit of account or a medium of exchange, respectively.

We have also presented a number of incidents that could be seen as a justification for further regulatory action.

---

[21] In a talk on 28C3, the 28th Chaos Communication Congress in Berlin, Germany, 2011, Dan Kaminsky describes an attack that involves connecting to all nodes of the Peer-to-Peer network; the first node announcing a transaction is considered as its source.